\documentclass{rspublic}

\usepackage{graphicx}
\usepackage{epsfig}

\def\spose#1{\hbox to 0pt{#1\hss}}

\def\kms{\ifmmode {\rm\,km\,s^{-1}}\else ${\rm\,km\,s^{-1}}$\fi}

\def\kmsmpc{\ifmmode {\rm\,km\,s^{-1}\,Mpc^{-1}}\else ${\rm\,km\,s^{-1}\,Mpc^{-1}}$\fi}

\def\ergps{\ifmmode {\rm\,erg\,s^{-1}}\else ${\rm\,erg\,s^{-1}}$\fi}
\def\ergpscm2{\ifmmode {\rm\,erg\,s^{-1}\,cm^{-2}}\else
    ${\rm\,erg\,s^{-1}\,cm^{-2}}$\fi}

\def\deg{\ifmmode {^{\circ}}\else {$^\circ$}\fi}
\def\degr{\ifmmode {^{\circ}}\else {$^\circ$}\fi}
\def\degs{\ifmmode {^{\circ}}\else {$^\circ$}\fi}

\def\etal{{\it et al.\ }}

\def\h3Mpc{h^{-3}{\rm Mpc}^3}

\def\arcsec{\ifmmode {^{\prime\prime}}\else $^{\prime\prime}$\fi}
\def\asec{\ifmmode {^{\prime\prime}}\else $^{\prime\prime}$\fi}
\def\arcmin{\ifmmode {^{\prime}}\else $^{\prime}$\fi}
\def\amin{\ifmmode {^{\prime}}\else $^{\prime}$\fi}

\def\secper{\ifmmode \rlap.{^{s}}\else $\rlap{.}{^{s}} $\fi}
\def\minper{\ifmmode \rlap.{^{m}}\else $\rlap{.}{^m} $\fi}
\def\secspt{\ifmmode \rlap.{^{\prime\prime}}\else
    $\rlap.{^{\prime\prime}}$\fi}
\def\arcsper{\ifmmode \rlap.{^{\prime\prime}}\else
    $\rlap.{^{\prime\prime}}$\fi}
\def\minspt{\ifmmode \rlap.{^{\prime}}\else
    $\rlap.{^{\prime}}$\fi}
\def\arcmper{\ifmmode \rlap.{^{\prime}}\else
    $\rlap.{^{\prime}}$\fi}
\def\spose#1{\hbox to 0pt{#1\hss}}
\def\simlt{\mathrel{\spose{\lower 3pt\hbox{$\mathchar"218$}}
     \raise 2.0pt\hbox{$\mathchar"13C$}}}
\def\simgt{\mathrel{\spose{\lower 3pt\hbox{$\mathchar"218$}}
     \raise 2.0pt\hbox{$\mathchar"13E$}}}

\def\refindent{\par\noindent\parskip=2pt\hangindent=3pc\hangafter=1 }

\def\aj{{\it AJ}}
\def\apj{{\it ApJ}}

\def\mnras{{\it MNRAS}}
\def\nature{{\it Nature}}

\def\ref#1;#2;#3;#4 {\refindent{#1,} {#2}, #3, #4}

\def\book#1;#2;#3 {\refindent{#1, }{in {\it{#2},} }{#3}}

\def\U300{\ifmmode{U_{300}}\else{$U_{300}$}\fi}
\def\B450{\ifmmode{B_{450}}\else{$B_{450}$}\fi}
\def\V606{\ifmmode{V_{606}}\else{$V_{606}$}\fi}
\def\I814{\ifmmode{I_{814}}\else{$I_{814}$}\fi}
\def\J110{\ifmmode{J_{110}}\else{$J_{110}$}\fi}
\def\H160{\ifmmode{H_{160}}\else{$H_{160}$}\fi}

\begin{document}

\title[The first galaxies]
{The first galaxies:  structure and stellar populations}

\author[M.\ Dickinson]{Mark Dickinson}
\affiliation{Space Telescope Science Institute, Baltimore, MD 21218, USA}

\label{firstpage}
\maketitle

\begin{abstract}{early universe --- galaxies: evolution --- 
galaxies: morphology -- galaxies: stellar content --- infrared: galaxies}

The Hubble Deep Fields continue to be a valuable resource for 
studying the distant universe, particularly at $z > 2$ where their 
comoving volume becomes large enough to encompass several hundred 
$L^\ast$ galaxies or their progenitors.  Here I present recent 
results from a near--infrared imaging survey of the HDF--North 
with NICMOS, which provides structural and photometric information 
in the optical rest frame ($\lambda\lambda_0$4000--5500\AA) for hundreds 
of ordinary galaxies at $2 < z < 3$, and which offers the means 
to search for still more distant objects at $z \gg 5$.  Lyman break 
galaxies at $2 < z < 3$ are compact and often irregular in the
NICMOS images;   ordinary Hubble sequence spirals and ellipticals 
seem to be largely absent at these redshifts, and apparently reached 
maturity at $1 < z < 2$.   The Lyman break galaxies have UV--optical 
spectral energy distributions like those of local starburst galaxies.
Population synthesis models suggest typical ages $\sim$few$\times 10^8$
years and moderate UV extinction ($\sim 1.2$~mag at 1700\AA), but the 
constraints are fairly weak and there may be considerable variety.  
Considering a near--IR selected galaxy sample, there is little 
evidence for a significant number of galaxies at $z \sim 3$ that 
have been missed by UV--based Lyman break selection.  Using the 
well--characterized $z \sim 3$ galaxy population as a point of 
reference, I consider Lyman break galaxy candidates at $4.5 < z < 9$,
as well as one remarkable object which might (or might not) be 
at $z > 12$.  The space density of UV--bright galaxies in the 
HDF appears to thin out toward larger redshifts, although surface 
brightness selection effects may play an important role.

\end{abstract}

\section{Introduction}

\subsection{The first galaxies?}

The past five years have seen remarkable breakthroughs in our ability
to identify and systematically study ordinary galaxies at very large 
redshifts, not just as isolated case studies, but {\it en masse} as 
a galaxy {\it population.}  To date, nearly 1000 galaxies have 
been spectroscopically confirmed at $z > 2$, mostly identified via 
broad--band color selection techniques (Steidel \etal 1996, i.e., 
the Lyman break galaxies, or LBGs), but with other important and 
complementary methods also coming into play (sub--mm and radio surveys, 
emission line searches, QSO absorption systems, etc.).  A broad census 
of galaxy properties at $z \sim 3$ now seems within reach, covering star 
formation rates (SFRs), dust content, morphologies, spatial clustering, 
and perhaps even chemical abundances and internal kinematics.

Although I have retained my assigned title, `The First Galaxies...,'
it is far from clear that we know when, where, or how to find the `first' 
galaxies.  The $z \sim 3$ LBGs may or may not be the first major wave 
of galaxy formation.  If anything, current data favors a roughly constant 
global SFR (as traced by cosmic UV luminosity density, at least) from 
$2 \simlt z \simlt 4$, with no certain evidence for a decline at higher 
redshifts (Steidel \etal 1999).  The sub--mm population detected by 
SCUBA (cf.\ Cowie, this volume) may or may not represent the bulk of early 
star formation, and the upper redshift bound to SCUBA sources remains unknown.  
The reionization of the intergalactic medium at $z > 5$ and the presence 
of metals in the Ly~$\alpha$ forest at $z \sim 3$ point toward earlier 
epochs of star and galaxy formation, at least in trace amounts.  
A handful of galaxies now have plausible spectroscopic confirmation 
at $z > 5$, but too few for any systematic census.   

For this reason, I cannot promise to live up to my title:  I do not know 
what the first galaxies are or what they look like.  Given this ignorance, 
I will focus the first part of this article on the structure and stellar 
populations of the most distant {\it well studied} galaxies, the $z \sim 3$ 
Lyman break objects.  This is not meant as a comprehensive review,
but will instead concentrate on new imaging and photometric data from 
NICMOS on {\it HST} that extend our knowledge of LBG properties to the optical 
rest frame.   In his contribution to this volume, Max Pettini provides 
a complementary discussion of recent efforts to measure chemical abundances 
and internal kinematics for these same galaxies.  In the second part of 
my article, I will describe efforts to extend Lyman break color selection 
to still larger redshifts, approaching (or perhaps even exceeding) 
$z \approx 10$.  In this way I hope to at least provide a look into 
the epoch when `the first galaxies' might plausibly have been formed, 
and to catalog what we can find right now, in the pre--{\it NGST} era, 
given the best available survey data.  

\subsection{Infrared observations of the Hubble Deep Field}

For the past five years, the Hubble Deep Fields (HDFs) have provided 
the most exquisitely deep, high angular resolution optical census of the 
distant universe.  It is important (if somewhat pedantic) to consider 
what an HDF is actually good for.  One WFPC2 field covers 5 arcmin$^2$, 
and probes a very small co--moving volume at $z < 1$, enough to hold only 
$\sim 12$--30 $L^\ast$ galaxies, depending on the cosmology.   Given small 
number statistics and concerns about clustering, the central HDF is therefore 
not the best place study massive galaxies in the `low' redshift universe, 
despite the fact that most of cosmic time and most bright galaxies with 
spectroscopic redshifts are at $z \simlt 1$.  There is far more volume at 
high redshift: 10.5 to 40$\times$ more at $2 < z < 10$ than at $z < 1$ for 
plausible cosmologies, room enough for several hundred $L^\ast$ galaxies 
or their progenitors.  

At $z > 1$, the optical light emitted from galaxies shifts into 
the near--infrared.  Thus in order to compare $z > 2$ galaxies to their 
local counterparts, and to search for still more distant objects at 
$z \gg 5$, it is important to extend the wavelength baseline.  The 
HDF--North was observed in the near--IR from the ground in several different 
programs (e.g., Hogg \etal 1997; Barger \etal 1998; and our own KPNO 4m 
$JHK_s$ imaging, cf.\ Dickinson 1998).  The depth and angular resolution 
(typically $\sim 1\arcsec$) of these data are a poor match to that of the 
optical WFPC2 HDF images.  Two programs therefore targeted the HDF--North
with NICMOS on board {\it HST}, providing much deeper images with high angular 
resolution.   The NICMOS GTOs (Thompson \etal 1999) imaged one 
NICMOS Camera 3 field ($\sim 51\arcsec\times 51\arcsec$) for 49 orbits 
each at F110W (1.1$\mu$m) and F160W (1.6$\mu$m).  Our own program 
mosaiced the complete HDF with a mean exposure time of 12600s per 
filter in F110W and F160W.  Sensitivity varies over the field of view, 
but the average depth is AB~$\approx 26.1$ at $S/N=10$ in an $0\secspt7$ 
diameter aperture.  The drizzled PSF has FWHM~=~$0\secspt22$, primarily 
limited by the NIC3 pixel scale.   Because most galaxies have spectral 
energy distributions (SEDs) which brighten (in $f_\nu$ units) at redder 
wavelengths, our images detect roughly half of the galaxies from the 
WFPC2 HDF,  despite their short exposure times.  We have also reanalyzed 
our KPNO $K_s$ images to optimally extract photometry matched to the 
WFPC2+NICMOS data.   These data are not as deep as one would like, 
which is unfortunate because they provide the only access to rest frame 
optical wavelengths for objects at $3 < z < 4.4$, but they are the best 
presently available.

Thanks to the dedicated efforts of the observers, a remarkably high 
density of spectroscopic redshifts is available in the HDF--N:  $\sim 150$ 
galaxies (plus a few stars) in the central WFPC2+NICMOS field 
alone, with 33 objects at $2 < z < 5.6$.  Taking advantage of the high 
quality photometric data, many investigators have used multicolor 
selection (e.g.\ the two color Lyman break criteria of 
Steidel \etal 1996, Madau \etal 1996, and others) or photometric 
redshifts (e.g., Fern\'andez--Soto \etal 1999) to identify high redshift 
galaxy candidates.  There are advantages and drawbacks to both approaches, 
but both have demonstrated remarkable successes.  In this discussion, 
I will make use of both methods.  For the photometric redshifts, 
I will use fits to our 7--band HDF photometry by Budav\'ari \etal (2000), 
whose `adaptive template' method is a modification of an otherwise 
straightforward spectral template fitting scheme.  I will use AB magnitudes 
here throughout, and notate the six WFPC2 + NICMOS bandpasses by 
\U300, \B450, \V606, \I814, \J110 and \H160.  Unless stated otherwise, 
I will assume a cosmology with $\Omega_M = 0.3$, $\Omega_\Lambda = 0.7$, 
and $H_0 = 70$~\kmsmpc.

\section{Galaxy morphologies at \mbox{\boldmath $2 < z < 3$}}

The NICMOS $H_{160}$ images sample rest frame wavelengths in the optical 
$V$ to $B$ bands from $z = 2$ to 2.8.  This upper bound is about the midpoint 
of the redshift range where \U300--dropout Lyman break selection in the 
HDF is most efficient.  Therefore, at these redshifts we may use the NICMOS 
data to study the morphologies of LBGs at wavelengths where long--lived 
stars, if they are present, may dominate the light from the galaxy, and 
where dust obscuration should play a significantly lesser role than it does 
in the ultraviolet.  For LBGs at $z \simgt 3$, the NICMOS $H_{160}$ bandpass 
slips into the rest frame ultraviolet, and the NICMOS images once again 
tell us more about the distribution of star formation within galaxies than 
about that of their stellar mass.

\begin{figure}
\centerline{\psfig{figure=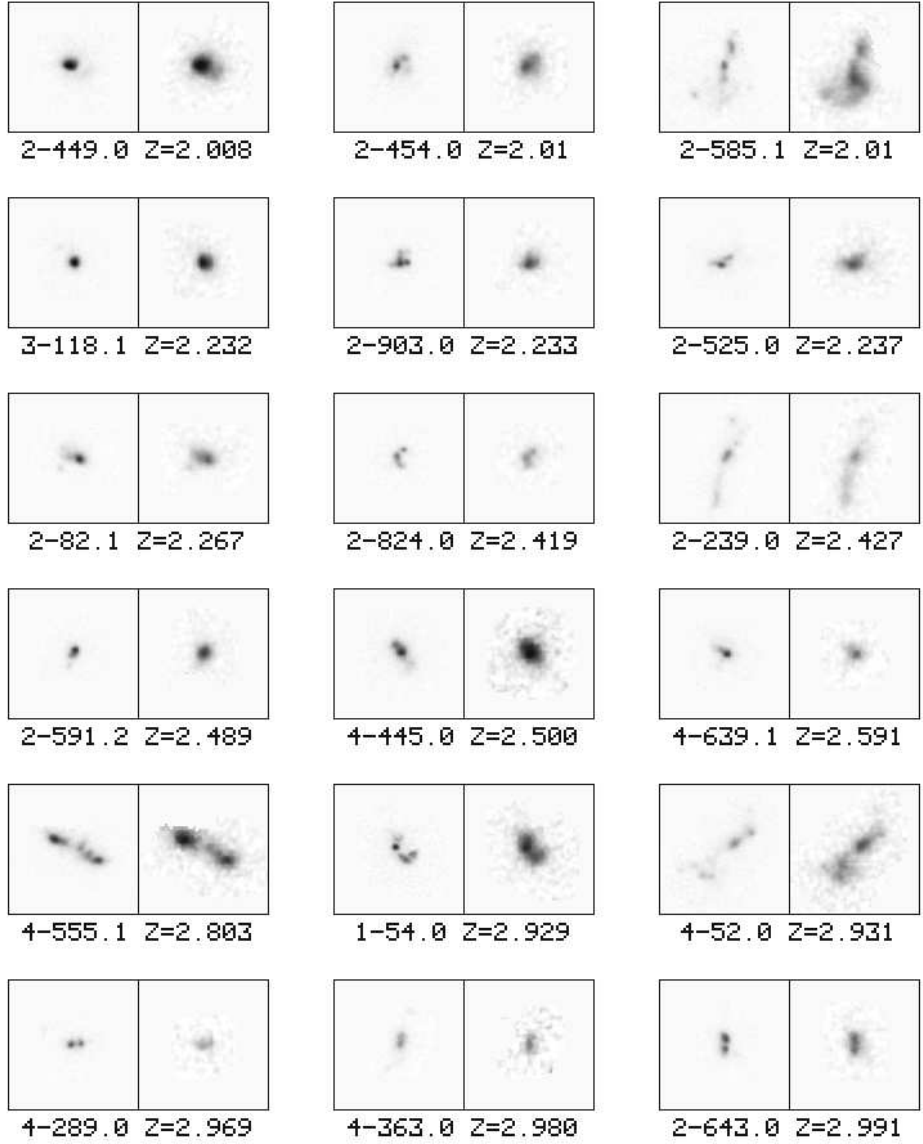,width=5in}}
\caption{
Ultraviolet and optical rest frame images of HDF Lyman break galaxies 
with spectroscopic redshifts $2 < z < 3$.  For each object, the left hand
image is interpolated between WFPC2 passbands to $\lambda_0$1700\AA, while 
the right panel interpolates the NICMOS images to the rest frame 
$B$--band (4300\AA).  Galaxies at $2.8 < z < 3$ use the NICMOS 
$H_{160}$ images without extrapolation.  Boxes are 
$4\arcsec \times 4\arcsec$, or 32$h_{70}^{-1}$~kpc on a side 
at $z=2.5$.  Here, the WFPC2 images have {\it not} been convolved 
to match the NICMOS PSF.
}
\end{figure}

Figure~1 compares rest frame UV and optical images of a set of HDF LBGs 
with spectroscopic redshifts $2 < z < 3$.  The NICMOS data have somewhat 
poorer angular resolution (0\secspt22\ compared to 0\secspt14\ for WFPC2), 
but otherwise the most striking thing is the broad similarity of the 
UV and optical morphologies.  Giavalisco \etal (1996) and 
Lowenthal \etal (1997) have emphasized the very small sizes of LBGs 
in WFPC2 images, and the same is true at NICMOS wavelengths.  
Accounting for PSF differences, the half--light radii of the galaxies 
measured from the NICMOS images are the same as (or in some cases, 
slightly smaller than) those from the WFPC2 data.  
Notable morphological differences are seen in only a few 
cases, e.g., 2--585.1 ($z = 2.008$) and 4--52 ($z = 2.931$), two of the 
largest LBGs in the HDF.  Each has a region of diffuse emission in the WFPC2 
images which lights up in the NICMOS $H_{160}$ data, with quite red 
colors.  It is not clear whether this is due to the presence of dust, 
older and redder starlight, or possibly strong line emission.\footnote{4--52 
is one of the few $z > 2$ galaxies detected at 8.5~GHz by Richards \etal 1998.
The radio centroid is coincident with the diffuse, red, IR--bright region 
in the galaxy.}

In general there is no evidence that the UV--bright regions seen by 
WFPC2 are just star--forming `fragments' embedded within some larger, 
mature host galaxy.  On the whole, one or more of the following appears
to be true:  (1) the stars which dominate the light at
$\lambda\lambda_0$1200--1800\AA\ also dominate at 
$\lambda\lambda_0$4000--5500\AA;  (2) if components with substantially 
different ages and colors are present, then they are fairly well mixed,
spatially; (3) dust extinction does not play a dominant role in shaping 
the morphologies of LBGs at these wavelengths.  All of this should be 
taken {\it modulo} the important caveat that at FWHM~$= 0\secspt22$ 
resolution, most of the LBGs are {\it not} exquisitely 
resolved:  they are typically very small, having only a few resolution 
elements within their isophotally detectable areas, and thus many details 
are surely lost except perhaps for the few, largest objects.

It is striking that among the NICMOS images of $\sim 100$ HDF LBGs (those
with and without spectroscopic redshifts), virtually {\it none} resembles 
a `classical' Hubble sequence spiral galaxy.  Giant disk galaxies are found 
in the HDF and elsewhere out to at least $z \approx 1.3$ (see, e.g., examples 
from NICMOS in Bunker 1999 and Dickinson 2000), and various studies have found 
that their structural properties and comoving abundances have not changed 
dramatically since $z \approx 1$ (Lilly \etal 1998;  Simard \etal 1999).
At $z > 2$, however, there appear to be no objects with immediately 
recognizable spiral structure, no bulges surrounded by symmetric, 
diffuse disks, nor even good candidates for thin, edge--on disks 
(some LBGs are fairly elongated, but none really could be mistaken 
for an edge--on spiral).   Among the objects shown in figure~1, 
perhaps 2--585.1 ($z=2.008$) appears closest to showing spiral structure 
at rest frame optical wavelengths, but this requires some imagination. 
Given the evidence from NICMOS, the absence of classical spiral 
morphologies cannot be attributed solely to rest frame wavelength 
effects.   We have carried out simulations using optical rest frame images 
of Virgo cluster spirals and of HDF disk galaxies at $z < 1$, artificially 
inserted into the NICMOS images at high redshift with the appropriate surface 
brightness dimming and PSF convolution.   Although low surface brightnesses
and limited angular resolution wipe out many of the details of spiral 
structure, giant ($L > L^\ast$) disk galaxies should be detectable and 
recognizable at $z > 2$ even with {\it no} luminosity or surface brightness 
evolution.  This is not to say that LBGs cannot be disk galaxies:  as already 
noted, the small angular sizes of most LBGs preclude detailed resolution, 
and some of these objects could well be small disks.   Indeed, 
Giavalisco \etal (1996) noted exponential surface brightness 
profiles among some LBGs.

Giavalisco \etal also found that some LBGs have $R^{1/4}$ law profiles, 
although it is not clear whether to interpret this as indicating a relation
between LBGs and present--day ellipticals and bulges.  It is worth noting, 
however, that there are few candidates for intrinsically {\it red} 
elliptical galaxies at $z > 2$ in the HDF, i.e., objects which would be 
recognized as elliptical galaxies today both by morphology and by the 
characteristic color signature of an older stellar population.  In fact, 
considering objects with spectroscopic or photometric redshifts in the 
range $2 < z < 4$, where the NICMOS+KPNO infrared data still reach the 
optical rest frame, there are only a handful of galaxies with rest frame 
colors redder than that of a present--day Scd spiral (i.e., an actively
star forming galaxy), even in an infrared--selected sample which should 
have no bias against such objects.   I return to this point in \S4 below.
Nearly every HDF galaxy at $z > 2$ which is detectable at $1.6\mu$m appears 
to be forming stars, and most quite vigorously.   We {\it do} find apparently 
red, dead ellipticals in the HDF out to (photometric) redshifts $z \approx 1.8$ 
(cf.\ Dickinson 2000;  Stanford \etal in prep.), but only one (marginally) 
viable candidate for a red elliptical at higher redshift.   This is the 
so--called `$J$--dropout' object HDFN--JD1 (see \S6), whose colors 
might be matched by those of a maximally old elliptical galaxy at 
$3 < z < 4$ (Dickinson \etal 2000).\footnote{In fact, HDFN--JD1 is probably 
too red for an old stellar population at high redshift without invoking 
dust, an unusual IMF, or an unfashionable cosmological model.}

Overall, it seems that the maturation of the giant spiral and elliptical
galaxies took place at $z < 2$.  Even with extremely deep, high angular 
resolution infrared images like the HDF/NICMOS data, we find few (if any) 
mature spirals or ellipticals (or candidates for such) at higher redshift.
By $z = 1$, many (if not necessarily all) large spirals and red giant 
ellipticals were already in place, pointing to the redshift range 
$1 < z < 2$ as an important `golden age' for the formation 
of the Hubble sequence.

\section{Galaxy colors at \mbox{\boldmath $2 < z < 3.5$}}

\begin{figure}
\centerline{\psfig{figure=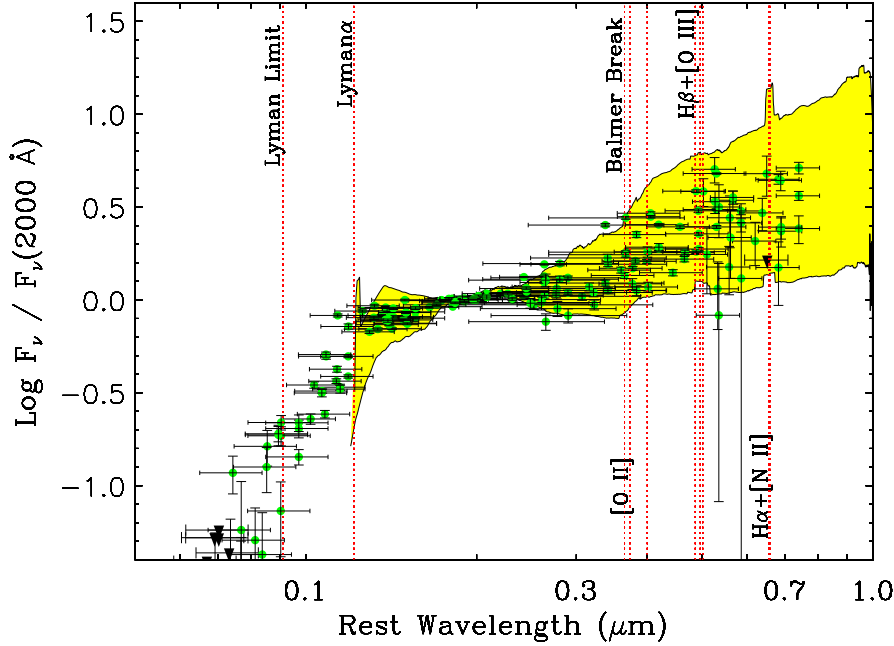,width=5in}}
\caption{Photometry for 27 HDF galaxies with spectroscopic redshifts
$2 < z < 3.5$, shifted to the rest frame and normalized at 2000\AA.  
The shaded region spans the range of empirical starburst templates 
from Kinney \etal (1996).  The starburst SED sequence is primarily 
defined by reddening up to $E(B-V) \simlt 0.7$;  the blue envelope 
is set by NGC1705.}
\end{figure}

Figure~2 shows a photometric compendium of spectroscopically 
confirmed HDF galaxies at $2 < z < 3.5$, all shifted to the rest 
frame and normalized to a common flux density at $\lambda_0$2000\AA.  
The shaded envelope is defined by local UV--to--optical starburst galaxy 
spectral templates from Kinney \etal (1996), which span a broad 
range in optical/UV extinction.  The HDF LBGs fall comfortably within
the range of SED shapes defined by the local starbursts.  They
have relatively blue (but usually not flat spectrum) UV continua, 
with a flux increase and spectral inflection around the Balmer/4000\AA\ 
break region that indicates the presence of older (A and later) stars
which apparently contribute a significant fraction of the rest frame 
optical light.  The fact that the UV continuum slope for LBGs is 
nearly always redder than flat spectrum has generally been interpreted 
as an indication of dust extinction (e.g., Meurer \etal 1997, 1999;  
Dickinson 1998;  Pettini \etal 1998), although for some objects it
might also result from an aging stellar population with declining or 
inactive star formation.   Considered individually, the large majority 
of HDF LBGs are reasonably well fit by the Kinney \etal starbursts 
with modest reddening ($0 < E(B-V) < 0.21$).  Few approach the more 
heavily reddened templates.  This may be true by definition/selection, 
of course, since we are considering the brighter objects for which 
redshifts were successfully measured, and which were selected for 
spectroscopy by their UV colors (but see \S4 below).

With photometry spanning the UV--to--optical rest frame it becomes 
interesting to compare the LBG photometry to population synthesis models 
to look for constraints on galaxy ages, reddening, and star formation 
histories.  Very roughly, if the UV spectral slope provides a measure of
extinction (modulo an assumed reddening law), then the UV--optical flux 
ratio, and particularly the amplitude of any inflection around the 
Balmer/4000\AA\ break region, may help constrain the past star formation 
history, particularly the ratio of older stars to ongoing star formation.  
This was first done by Sawicki \& Yee (1998) using ground--based 
$JHK_s$ photometry for LBGs in the HDF.  Their results favored young 
ages (median value $\sim 25$~Myr) and fairly heavy reddening (typical 
$E(B-V) \approx 0.28$, or 3~mag extinction at 1600\AA\ assuming 
Calzetti 1997 starburst dust attenuation).  We have carried out 
a similar exercise using the NICMOS data and $K$--band fluxes rederived 
from the KPNO data using a technique (much like that of 
Fern\'andez--Soto \etal 1999) which properly matches photometry 
from images with very different angular resolutions.   A complete 
presentation will be given in Papovich \etal (in prep.) and is 
beyond the scope of the present discussion, but I summarize some 
important points here.

\begin{figure}
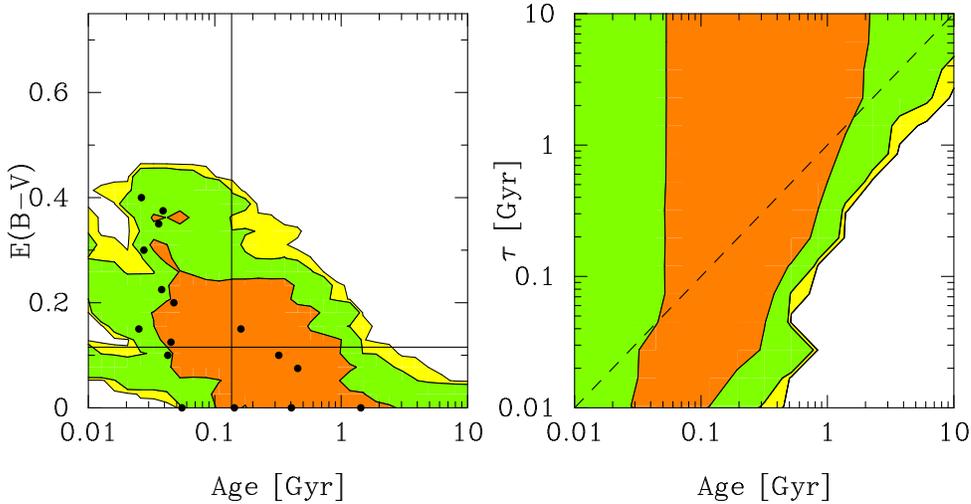

\centerline{\psfig{figure=lbg_ebvage_zlo.ps,width=2.5in} 
	\psfig{figure=lbg_tauage_zlo.ps,width=2.5in}}
\caption{
Combined confidence intervals (68\%, 95\%, 99.7\%) for population
synthesis model fits to a sample of 16 HDF Lyman break galaxies with 
spectroscopic redshifts $2 < z < 2.95$, from Papovich \etal (in prep.).  
A wide range of Bruzual \& Charlot (1996) models has been considered with 
varying age ($t$), star formation histories (exponential timescale $\tau$), 
dust attenuation ($E(B-V)$, here assuming the Calzetti 1997 starburst law), 
metallicity, and IMF.   Here, results for the 16 objects have been averaged, 
weighted by their individual probability distributions, to show the ensemble 
likelihood distributions for $E(B-V)$, $t$ and $\tau$.  Best--fitting values
for each object are marked by points in the left panel.  Dependences 
on metallicity and IMF have been collapsed, but are fairly weak.  
The cross--hairs at left indicate the most--favored values for 
$E(B-V)$ vs.\ $t$.  The dashed line at right marks $t = \tau$.  
Models observed after the bulk of their star formation has been 
completed (i.e.\ $t \gg \tau$), are disfavored except at young 
ages ($\simlt$~few$\times 10^8$~years) and short timescales, 
when the UV continuum can persist throughout the main sequence 
lifetimes of B stars.
}
\end{figure}

Even with precise NICMOS photometry, constraints on ages and reddening 
are quite loose.  This is in part because of the usual degeneracies in 
fitting models to broad band colors (age vs.\ metallicity vs.\ extinction), 
but also because the available photometry simply does not reach long enough 
rest frame wavelengths.   At $z > 3$, only the lower--$S/N$ ground--based 
$K_s$ data extend redward of the Balmer/4000\AA\ break region;  for
this reason, here I consider only galaxies at $2 < z < 3$.  For each 
galaxy, we may define confidence intervals in the multidimensional space 
of the various fitting parameters such as age, SFR e--folding timescale, 
reddening, metallicity, and IMF.  The color degeneracies allow fairly 
broad ranges in acceptable parameters for each object, and the variety 
among the galaxy SEDs (figure~2) scatters the best--fitting parameters 
throughout a range of values.  Nevertheless, taken as an ensemble, 
certain regions of model parameter space are preferred.   

Figure~3 shows a composite distribution for 16 galaxies with $z < 3$, 
where model fits for each object have been averaged, weighted by their 
likelihoods in the multi--parameter space.   The contours thus indicate 
a distribution of likely parameter values for the ensemble of UV--selected 
LBGs.  The best--fitting values of $E(B-V)$ vs.\ age for individual objects
are marked by dots.   Some galaxies can be fit reasonably well by parameter 
values falling toward the outer contours of the ensemble distribution, but 
the majority occupy the higher confidence regions.  The most favored age range 
spans 0.03--1~Gyr, with extinction $0 < E(B-V) < 0.25$.  The extinction
values agree well with the comparison to the Kinney \etal starburst 
templates (figure~2), which is not unexpected given that the Calzetti
attenuation law is derived in part from the same UV spectral data 
on local starburst galaxies.    The range of likely age and extinction
values becomes slightly smaller and can shift somewhat if restrictions 
on metallicity or IMF are adopted, although in general the broad band 
photometry offers little constraint on these parameters.  Younger ages 
and larger extinction are allowed and even favored for some objects, 
but in general the most likely values are somewhat older and less reddened 
than those found by Sawicki \& Yee.  This may be due to more precise 
photometry from NICMOS, or to better control of the relative optical--IR 
colors (particularly in the $K$--band) from our photometric method.  

It is not clear whether the apparent anticorrelation between age and 
extinction in figure~3 is significant.  We expect degeneracy between 
the age and extinction values fit to individual objects, but the overall 
trend seen in the population may be greater than would be expected from 
the fitting uncertainties alone.   The `most favored' extinction value, 
$E(B-V) \approx 0.12$, corresponds to $A(1700{\rm \AA}) \approx 1.2$~mag,
or a factor of $\sim 3$.  The {\it net} UV extinction for the sample 
(and hence the correction to any derived global star formation rate) 
would be larger, however, driven by the objects with the greatest reddening.

A characteristic timescale for Lyman break galaxies can be defined from 
their sizes (median half--light radii $\approx 2.2$~kpc for the HDF LBGs) 
and typical velocity dispersions ($\sim 80$~km~s$^{-1}$; see Pettini, this 
volume).  This yields $t_c \sim 25$~Myr.  If the UV light is due to 
ongoing star formation, we would not expect SFR lifetimes $\ll t_c$, 
and indeed this is roughly the lower bound of the best--fit model age 
range, with the most favored value being $\sim 135$~Myr, or $\sim 5t_c$.
This range is not dissimilar to that estimated for star formation 
in galactic--scale starburst events (e.g., ULIRGs) locally.

\section{Rest frame UV selection:  what do we miss?}

Using an infrared--selected catalog, we may ask what galaxies 
might be missed altogether by Lyman break color selection keyed to 
the rest frame UV light.   In particular, one might expect some 
red high redshift galaxies, either because they are not actively 
forming stars or because of extinction, that would `drop out' of 
the dropout samples.  Here I restrict my analysis to $H_{160} < 26$, 
where we believe our catalogs are highly complete, uncontaminated by 
spurious sources, and where the NICMOS photometry has $S/N \simgt 10$.
At $z = 2.75$, $H_{160} < 26$ corresponds to rest frame 
$M_B < -19.36$ for the adopted cosmology, or $\sim 1$~mag fainter
than present--day $L_B^\ast$.  
The typical LBG at $H_{160} \approx 26$ has $V_{606} \approx 27$,
the practical limit for HDF \U300--dropout selection using standard 
2--color criteria, but red galaxies with similar rest frame optical 
luminosities might be fainter or absent in the UV.   

At $H_{160} < 26$ there is only one object which is undetected with 
$S/N < 2$ in \V606 or \I814 (both, in this case):  this is the
`$J$--dropout' HDFN--JD1 (see also \S\S2 and 6).  In fact, this 
is the only NICMOS--selected object with $\H160 < 26$ 
and $S/N(\I814) < 6.5$.  Two other objects have $S/N(\V606) < 3$; 
both are $z \simgt 5$ \V606--dropout candidates identified by 
Lanzetta \etal (1996) and Fern\'andez--Soto \etal (1999), 
one of which (3--951) was spectroscopically confirmed at $z=5.33$ 
(Spinrad \etal 1999).   Thus the only possible candidate for 
a NICMOS--selected, `UV--invisible' galaxy at $z \sim 3$ is HDFN--JD1.  

\begin{figure}
\centerline{\psfig{figure=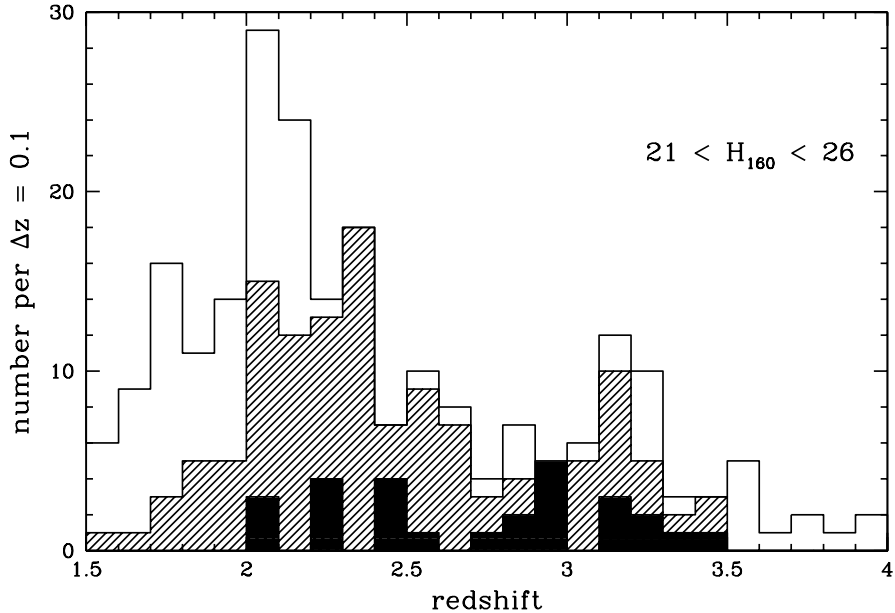,width=5in}}
\caption{Photometric and spectroscopic redshift distributions for
infrared--selected HDF galaxies ($H_{160} < 26$) at $1.5 < z < 4$.  
The open histogram shows the $z_{phot}$ distribution from 
Budav\'ari \etal (2000).  The hatched histogram indicates galaxies 
which obey the Lyman break criteria defined in the text, while 
the filled histogram shows the available spectroscopic redshifts.}
\end{figure}

Next let us consider UV--bright objects which might nevertheless
have been missed by the LBG color criteria, using the 7--band photometric 
redshift estimates for all galaxies.  In principle, these may identify 
plausible candidates at $2 < z < 3.5$ that otherwise fall outside a given 
set of UV color criteria, as long as their intrinsic SEDs are `recognizably 
similar' to those of galaxies at lower redshift which define the templates 
used for the phot--$z$ fitting.  

Figure~4 shows the $z_{phot}$ distributions for all HDF galaxies 
with $21 < H_{160} < 26$, and for those that meet the \U300 dropout 
criteria $(\U300 - \B450) > (\B450 - \V606) + 1$ and $\B450 - \V606 < 1.2$ 
from Dickinson (1998).  There are 43 objects with $2 < z_{phot} < 3.5$ which 
do {\it not} meet the LBG color criteria.  However, nearly all are at 
$2 < z_{phot} < 2.2$ or $3.2 < z_{phot} < 3.5$, and lie just outside 
the color selection boundaries defined here: either slightly 
too blue at low $z$, or slightly too red at high $z$.  This is 
expected:  the selection efficiency of the 2--color method
is not uniform with redshift, and falls off at the extremes of
the range for which it is optimized (cf.\ Steidel \etal 1999).
Only 7 `missed' objects fall at intermediate photometric redshifts, 
$2.5 < z_{phot} < 3.1$, and most of these are also just outside
the color selection box.  Some are quite interesting, including
a $\mu$Jy radio source with very red $\J110 - \H160$ colors which 
may be a dusty starburst or a fading post--starburst galaxy 
at $z \sim 2.6$.  Others may be scattered out of the box by 
photometry errors (especially in \U300), or might not be at the 
indicated $z_{phot}$.  But all are well detected in the optical HDF.

Overall, there is no evidence for a {\it substantial} population 
(by number) of galaxies at $2 < z < 3.5$ that are missed by UV Lyman 
break color selection but which are detectable in the near--infrared. 
If there are energetically important but highly obscured galaxies at 
these redshifts like those detected by SCUBA, then they are either 
{\it also} detectable with optical imaging data, or they are so 
heavily enshrouded that even NICMOS cannot easily see them.  For 
the five sub--mm sources detected in the HDF by Hughes \etal (1998), 
our NICMOS images do not reveal any new counterparts previously 
undetected by WFPC2, nor do any of the candidate identifications 
have particularly unusual optical--IR colors.

\section{Galaxies at \mbox{\boldmath $4.5 \simlt z \simlt 9$}}

\begin{figure}
\centerline{\psfig{figure=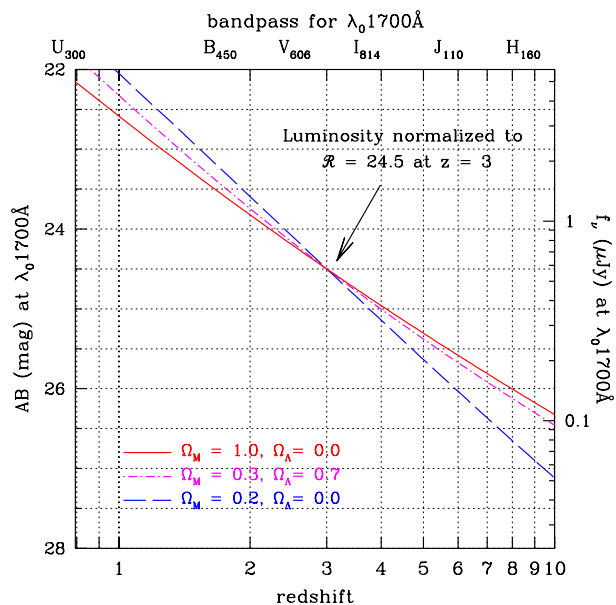,width=3.5in}}
\caption{Apparent magnitude for an $L^\ast$ ($\lambda_0 1700$\AA\
at $z \approx 3$;  Steidel \etal 1999) Lyman break galaxy
redshifted without evolution.  No $k$--correction is 
included:  instead, the detection bandpass is taken to be
fixed to $\lambda_0 1700$\AA\ and changes with redshift 
(see top axis labels).  An $L^\ast$ LBG should be detectable 
in the NICMOS data ($H_{160} < 26.5$) out to $z \approx 10$ for 
spatially flat cosmologies with $\Lambda < 0.8$, and to 
$z \approx 7.5$ for an open universe.
}
\end{figure}

\begin{figure}
\centerline{\psfig{figure=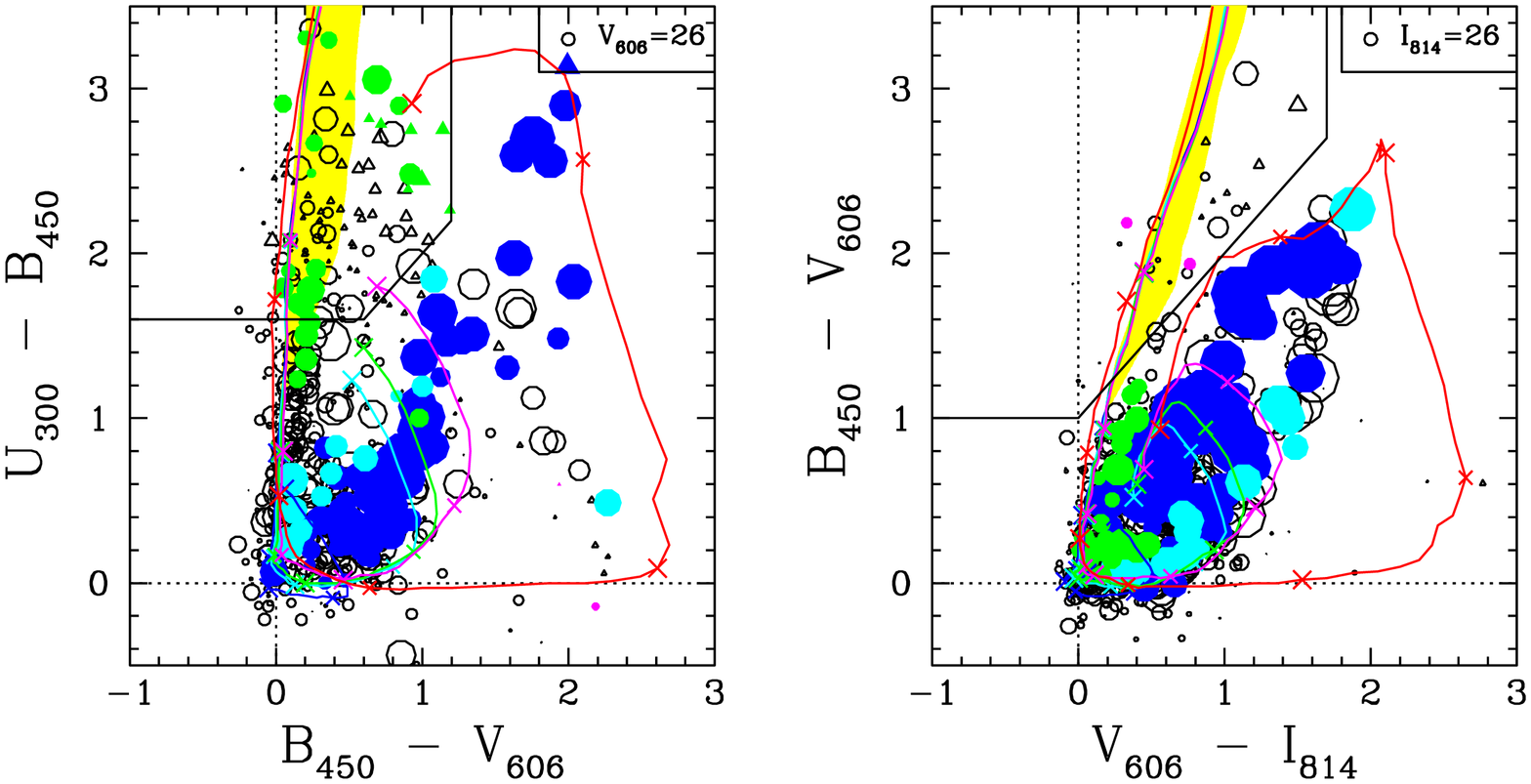,angle=-90,width=5in}}
\centerline{\psfig{figure=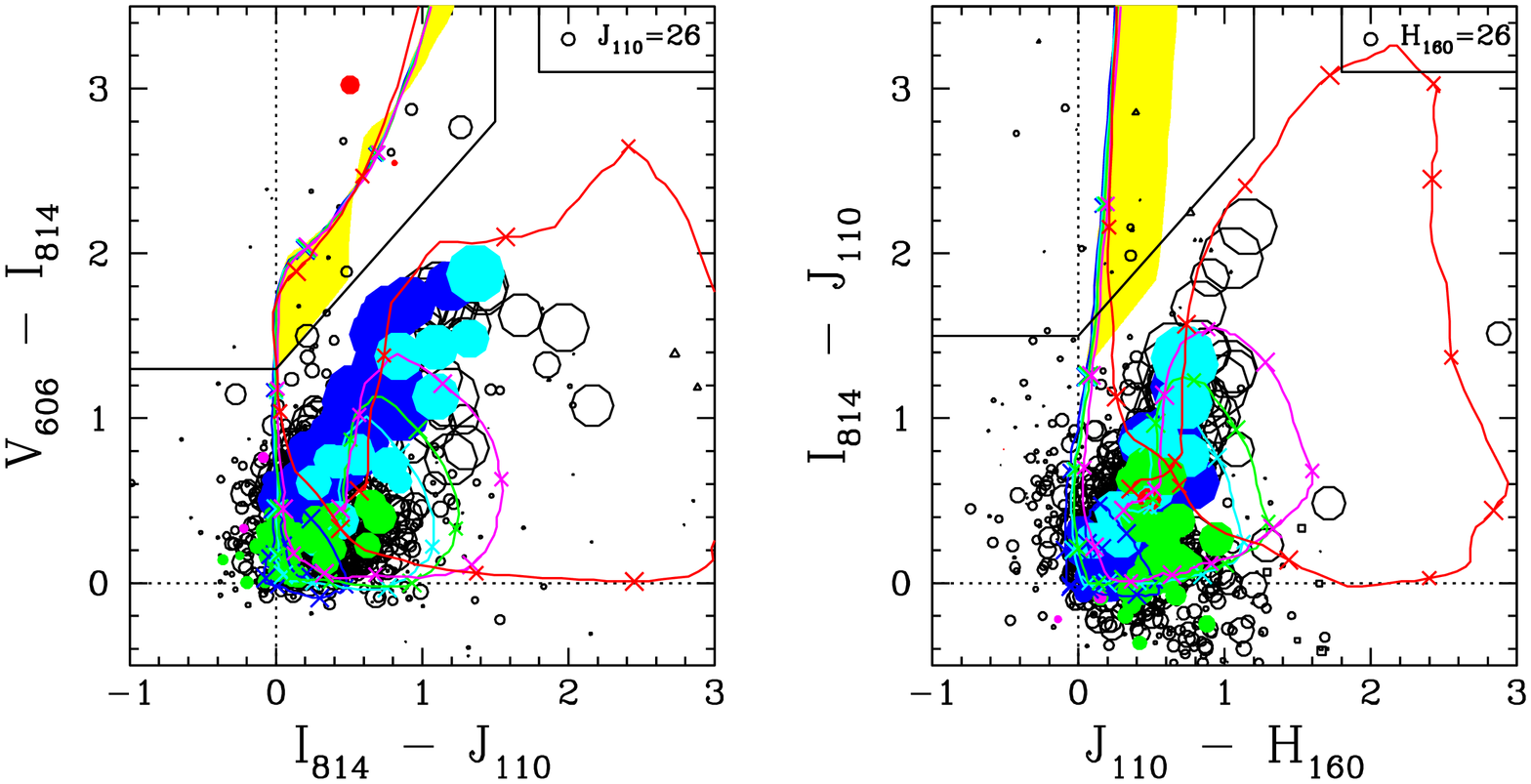,angle=-90,width=5in}}
\caption{Two--color diagrams for combinations of adjacent
HDF WFPC2+NICMOS filters.  Symbol size scales with magnitude;  
known and suspected stars and a few bright galaxies have been 
excluded for clarity.  Triangles mark $1\sigma$ lower color 
limits;   filled symbols indicate galaxies with spectroscopic
redshifts.  The lines show the nominal color--vs--$z$
tracks for various unevolving galaxy SEDs, and the shaded
region indicates the approximate color range expected
for galaxies in redshift ranges appropriate to each color
pair ($UBV$: 2--3.5;  $BVI$:  3.5--4.5;  $VIJ$:  4.5--6;
$IJH$:  6--8.5).  The color selection boxes used for the 
comparison with Monte Carlo simulations are indicated.}
\end{figure}

\begin{figure}
\centerline{\psfig{figure=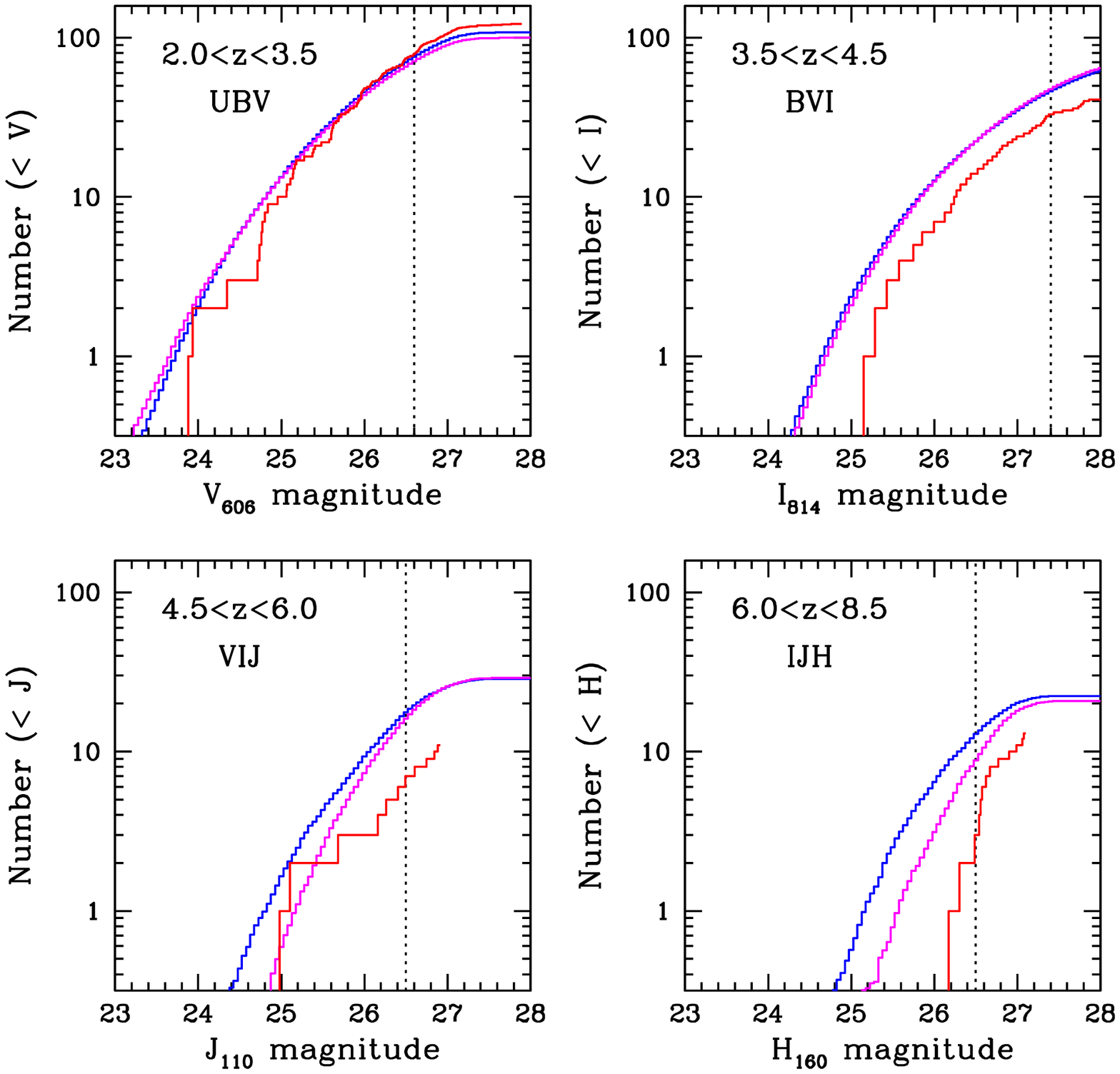,width=5in}}
\caption{Cumulative number counts of objects satisfying
the high redshift color criteria shown in figure~6.
The irregular histograms are from the HDF data, and the 
smoother histograms are predictions from the Monte Carlo 
simulations which assume that the galaxy population at high 
redshift is like that observed at $z \approx 3$.   Models were 
computed for two $\Lambda = 0$ cosmologies:  $\Omega_M = 1.0$ 
and 0.2 (the open cosmology is the lower model in the $VIJ$
and $IJH$ plots).  The vertical dashed line indicates a rough 
confidence limit in magnitude for each plot, below which 
the data should become significantly incomplete and/or 
contaminated by spurious sources.}
\end{figure}

The successes of color selection techniques at $2 \simlt z \simlt 4.5$ 
make it tempting to extend the methods to higher redshifts, i.e., 
to search for $V$-- or $I$--dropouts.  Doing so properly requires 
deep near--infrared data to provide at least one color longward 
of the redshifted 912\AA\ and 1216\AA\ breaks.  Indeed, galaxies at 
$z > 6.5$ should have virtually no detectable optical flux.  
Lanzetta \etal (1996) and Fern\'andez--Soto \etal (1999) 
identified candidate $z \simgt 5$ HDF galaxies from $\V606 - \I814$ 
colors supplemented by infrared limits from the KPNO data; two of these 
have subsequently been confirmed via spectroscopy (Weymann \etal 1999; 
Spinrad \etal 1999).  With NICMOS we can extend this to fainter 
limits and larger redshifts:  an $L^\ast$ Lyman break galaxy 
(i.e., $L^\ast$ in the rest frame UV at $z = 3$) should be detectable 
in the NICMOS images with $H_{160} < 26.5$ out to $z=10$ for spatially 
flat cosmologies with $\Lambda \leq 0.8$, and out to $z \approx 7.5$ for 
an $\Omega_M = 0.2$ open universe (see figure~5).

Here we test the null hypothesis that the galaxy population at $z \gg 3$ 
is similar to that of the \U300--dropout LBGs at $z \sim 3$, whose 
{\it observed} characteristics are by now reasonably well known even 
if their intrinsic properties, such as dust content, star formation rate, 
mass, etc.\ are the subject of continued debate.  In particular, we adopt 
the rest frame UV luminosity function and UV spectral slope (i.e., 
intrinsic color) distribution for LBGs at $\langle z \rangle \approx 3$ 
derived in Steidel \etal (1999), and use this to predict what should 
be seen in color--color diagrams {\it if\ } the same population were present 
at higher redshifts.  We do this via Monte Carlo simulations, including 
realistic errors for the HDF WFPC2+NICMOS photometry, comparing the number 
of high--$z$ objects that are predicted to fall in some specified 
color--color box to the actual number of similar objects found 
in the HDF catalogs.  

Figure~6 shows a series of 2--color diagrams for the HDF, 
each using combinations of three adjacent bandpasses, from 
$UBV$ (i.e., $z \sim 3$ selection) through $IJH$ (i.e., 
$z \sim 7$).  In each case, I define somewhat arbitrary 
selection boxes based on the expected location of high 
redshift galaxies in color--color space (and also to avoid 
low redshift contaminants), then count the galaxies in those 
boxes and compare this to the `no evolution' (NE) model 
predictions (figure 7).

\begin{figure}
\centerline{\psfig{figure=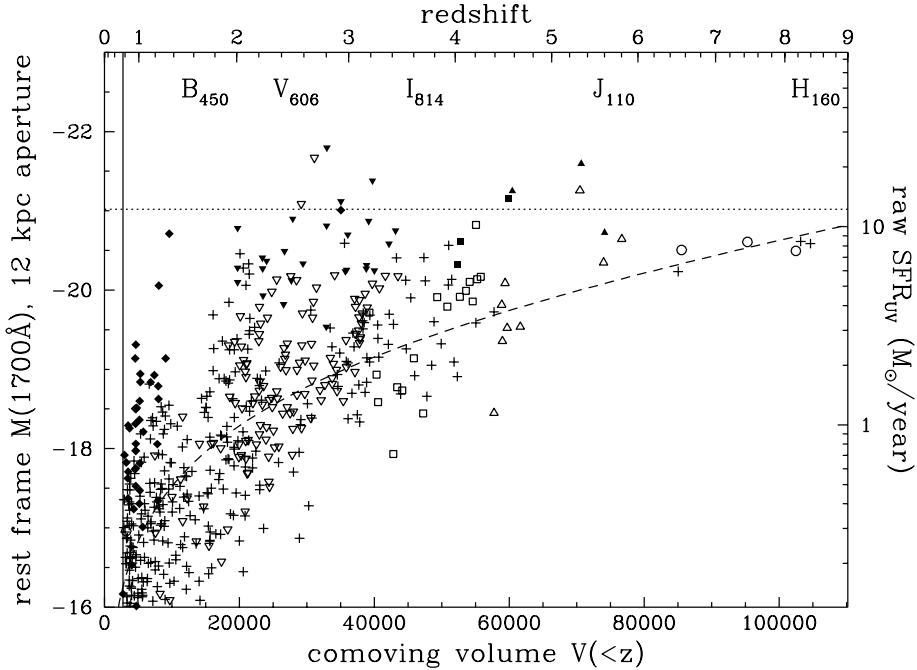,angle=-90,width=5in}}
\caption{
UV luminosity at $\lambda_0 1700$\AA\ vs. co--moving volume out
to redshift $z$ for galaxies with $H_{160} < 26.5$.  A constant 
horizontal density of objects implies a constant space density. 
Fluxes are measured through 12$h_{70}^{-1}$~kpc diameter metric 
apertures at all redshifts.  The right--hand axis indicates the 
corresponding `raw' star formation rate derived from the UV 
luminosity without correction for dust, assuming a Salpeter IMF.
The horizontal dotted line marks the luminosity of an $L^\ast$ LBG 
at $z \approx 3$ (Steidel \etal 1999).  The dashed curve indicates 
$m(\lambda_0 1700) = 26.5$ at the bandpass indicated by labels 
at top.  Objects with spectroscopic redshifts 
are indicated with filled symbols;  photometric redshifts are used 
otherwise.  Objects meeting 2--color criteria as $\U300$, $\B450$, 
$\V606$ and $\I814$ dropouts are coded as $\bigtriangledown$, 
$\square$, $\bigtriangleup$, and $\bigcirc$, respectively;  
all others are plotted with $+$ and $\diamond$.  
}
\end{figure}

The $\U300$--dropout counts agree well with the models by construction, 
since the input luminosity function is partially based on HDF data.  
The $\B450$--dropouts fall below the NE predictions.  This is just 
the original Madau \etal (1996) result revisited:  the HDF--N 
appears to have fewer galaxies at $z \sim 4$ than at $z \sim 3$.  
Steidel \etal (1999), who surveyed larger solid angles in several 
fields, suggest that the bright end of the $z \sim 4$ LF is actually 
compatible with that at $z \sim 3$.  The HDF may just be an anomaly, 
indicating the importance of field--to--field fluctuations, or perhaps 
the faint end slope of the LF (to which the HDF number counts are
quite sensitive) evolves with redshift.   For the $\V606$--dropouts,
there are $\sim 7$ candidates with $\J110 < 26.5$ (including the 
two with spectroscopic confirmation), compared to a prediction of 
$\sim 17$.  Careful inspection of their images and SEDs suggests that 
they are all very plausible $4.5 \simlt z \simlt 6$ candidates.   
There are {\it no} $\I814$--dropout candidates with $\H160 < 26$, and 
only two with $\H160 < 26.5$, one of which is clearly detected at $\B450$ 
and $\V606$ and thus is probably not at $z > 6$.  The models predict 
9 to 13 objects to this magnitude limit.  Some of the fainter objects 
may be real $z > 6$ galaxies, but on visual inspection many are rather 
dubious, with very low $S/N$;  only a few are persuasive to a skeptical 
eye.   At $H_{160} > 26.5$ we are reaching or passing the useful depth 
limits of our NICMOS data for this purpose.

This analysis is in qualitative agreement with one based on 
photometric redshift estimates.  Figure~8 plots rest frame
1700\AA\ luminosities of galaxies vs.\ redshift (spectroscopic
when available, photometric otherwise) for an HDF sample limited
to $H_{160} < 26.5$.  The photometric redshifts are generally
in good agreement with the simple 2--color selection illustrated
in figure~6.  There are a few objects with $z_{phot} \approx 3.5$
that `fall in the gap' between the $U_{300}$-- and $B_{450}$--dropout
samples.  At $z_{phot} > 6$ there is only partial overlap between
the $I_{814}$--dropout and photometric redshift samples, but this
is understandable since most of the candidates are very faint 
with low--$S/N$ photometry and poor photometric redshift constraints
(i.e., relatively flat $z_{phot}$ likelihood functions).  The space 
density of the bright LBG candidates appears to thin at $z \simgt 4.5$, 
and at $z > 5.5$ there are no candidates with UV luminosities greater 
than the characteristic $L^\ast$ at $z = 3$, despite abundant volume 
to house them if they were present with similar space 
densities.\footnote{For an open universe, the higher $z_{phot}$ 
candidates would be more luminous and the upper envelope to their 
UV luminosity would be nearly flat, but the space densities would 
be even more sparse compared to those at $2 < z < 3.5$.}

Ferguson (1998) and Lanzetta \etal (1999) have stressed the importance 
of cosmological surface brightness dimming when characterizing the galaxy 
population at $z > 3$.  This can affect the likelihood of detecting 
high redshift galaxies, as well as the fluxes measured 
with isophotally--based photometry schemes.  Lanzetta \etal show that 
the global rate of star formation occurring in the regions with the 
highest UV surface brightness rises steeply with redshift, and argue 
that far more UV light may be present in $z > 5$ galaxies at fainter,
unmeasured isophotal thresholds.  We may partially address this by 
measuring fluxes and luminosities non--isophotally, e.g. by using 
apertures scaled by image moments or with fixed metric sizes.  The 
photometry in figure~8 uses 12~kpc metric apertures, and thus is 
insensitive to surface brightness limits except as far as they affect 
galaxy detection in the first place:  most faint object cataloging packages 
use isophotal detection thresholds.   We have examined this by taking 
WFPC2 images of $z \sim 2.5$ LBGs, artificially shifting them to higher 
redshifts, and re--inserting them into the NICMOS data to assess their 
detectability.  Galaxies like the $L \simgt L^\ast$ LBGs at $z \sim 3$ 
should be detectable to at least $z \approx 7$ at the depth of our 
GO NICMOS images, and more easily in the HDF--South NICMOS field or the 
HDF--North GTO NICMOS image, each of which goes $\sim 1$ magnitude deeper 
than the data set discussed here.  But a general census at such redshifts 
might indeed be woefully incomplete.

Overall, the HDF data disfavor the null hypothesis that galaxies 
like the bright LBGs at $2 < z < 4.5$ are present at $z \gg 5$ 
with similar space densities.  The higher redshift galaxies are 
apparently either fainter, more rare, have lower surface brightness, 
or some combination thereof.   At any rate, they are certainly 
more difficult to detect and study, at least in abundance, 
even with NICMOS.

\section{An object at \mbox{\boldmath $z \simgt 12$}?}

Although the few $I$--dropout candidates in the HDF are very
faint, paradoxically there is one comparatively bright `$J$--dropout' 
object, shown in figure~9.  Lanzetta \etal (1998) identified five 
possible sources in the KPNO $K_s$ images of the HDF which were 
invisible in the WFPC2 data.  Of these, four are undetected by NICMOS.  
One, however, which we call HDFN--JD1, has a robust 1.6$\mu$m 
detection ($\H160 \approx 25.2$), but is at best only marginally 
($S/N < 2$) detected at 1.1$\mu$m and shorter wavelengths.  
This optical `detection,' if real, would be important, 
as it would probably exclude the most exotic hypothesis 
for this object, i.e., that it is a galaxy or QSO at $z \simgt 10$.  
It will be difficult, however, to obtain much deeper optical data 
than the existing HDF WFPC2 images to provide a stricter limit.
The red $\H160 - K_s$ color suggests $z \approx 12.5$ under 
the high redshift hypothesis, with the Ly$\alpha$ forest 
partially suppressing the \H160 flux.  We obtained an $H$--band 
spectrogram of HDFN--JD1 with CRSP at the KPNO 4m, and to our 
surprise detected a moderately convincing emission line at 1.65$\mu$m 
that could plausibly agree with Ly$\alpha$ at $z = 12.5$.  
The line did not reproduce, however, in a subsequent 
reobservation at higher dispersion (see Dickinson \etal 2000 
for the spectra and further discussion). 

\begin{figure}
\centerline{\psfig{figure=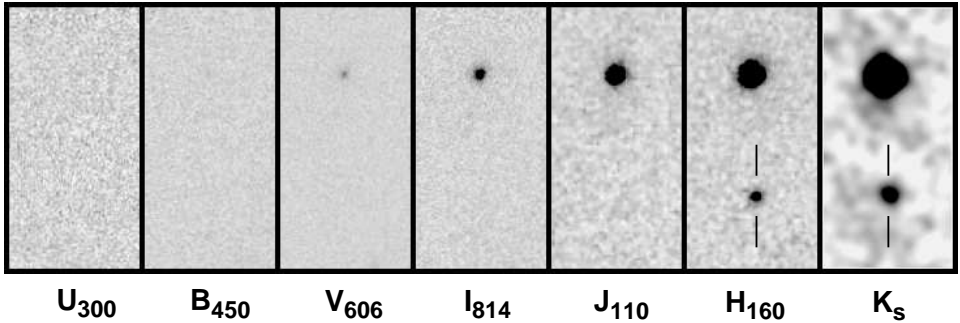,angle=-90,width=5in}}
\caption{
{\it HST} and Keck images of HDFN--JD1 at 0.3--2.16$\mu$m.  
The field of view of each panel is $4\arcsec \times 8\arcsec$.  
HDFN--JD1 is identified by tick marks in the \H160 and $K_s$ 
panels.  The Keck $K_s$ image has been smoothed by a Gaussian 
with FWHM~$= 0\secspt38$.
}
\end{figure}

If this is not a Lyman break object then it may be either heavily
reddened and at arbitrary redshift (but most likely $z > 2$, given the 
colors), or possibly a maximally old elliptical galaxy at
$3 < z < 4$ (see \S2).  If it is really at $z \approx 12.5$, then it 
is either a galaxy whose unobscured star formation rate (computed from 
the UV luminosity) is several hundred $M_\odot$~yr$^{-1}$, or an AGN, 
perhaps one of the hypothesized population responsible for re--ionizing 
the universe.  If so, however, such objects are rare at $2 < z < 13$, 
with a space density several hundred times lower than that of 
present--day $L^\ast$ galaxies, and it is unlikely that most of 
today's galaxies began their life in such a way.

Table 1 compares the UV luminosities of some confirmed and candidate
galaxies at $z > 4$, compared to that of an $L^\ast$ Lyman break galaxy
at $z = 3$.  In some cases infrared data needed to measure fluxes at 
rest frame 1700\AA\ are not available, and estimates based on published 
photometric or spectroscopic fluxes redward of Lyman~$\alpha$ have been 
used instead.\footnote{$L_\nu$(1700\AA) estimates for RD1 are derived 
in part from a $J$--band measurement by Armus \etal (1998).  For HDF 
3--951 and 4-473 we use our own NICMOS $J_{110}$ photometry.}
`Raw' star formation rates are computed from the 
UV luminosities assuming a Salpeter IMF and no dust obscuration.

\begin{table}
\caption{Properties of selected high redshift galaxies and galaxy candidates}
\longcaption{
Assumes $\Omega_M = 0.3$, 
$\Omega_\Lambda = 0.7$, $h = 0.7$, and that SFR = $1 M_\odot$~year$^{-1}$ 
produces $L_\nu (1700$\AA$) = 9 \times 10^{27}$ erg~s$^{-1}$~Hz$^{-1}$.  
Magnitudes are on the AB scale. 
}
\begin{tabular}{lllllll}
\hline
name & \multicolumn{1}{c}{z} & \multicolumn{1}{c}{$m$} &
\multicolumn{1}{c}{$M$} & \multicolumn{1}{c}{$L/L^\ast$} & 
\multicolumn{2}{l}{`raw' SFR ~~~~ reference} \\
 &   & \multicolumn{3}{c}{(UV at $\sim \lambda_0 1700$\AA)} 
 & \multicolumn{2}{l}{($M_\odot$~yr$^{-1}$)}  \\
\hline
$L^\ast$ LBG & ~~3.04   & 24.5  & -21.1  & $\equiv 1$ & ~12.8  & Steidel \etal 1999 \\
CDFa--G1     & ~~4.82   & 23.6  & -22.8  &  ~~4.7     & ~60    & Steidel \etal 1999 \\
RD1          & ~~5.34   & 25.5: & -21.0: &  ~~0.9:    & ~12:   & Dey \etal 1998 \\
HDF~3-951    & ~~5.33   & 25.0  & -21.5  &  ~~1.5     & ~19.6  & Spinrad \etal 1998 \\
HDF~4-473    & ~~5.60   & 26.0  & -20.6  &  ~~0.66    & ~~~8.5 & Weymann \etal 1998 \\
HCM1         & ~~5.74   & 25.5: & -21.1: &  ~~1.1:    & ~13.5  & Hu, Cowie \& McMahon 1999 \\
A            & ~~6.68   & 23.9: & -23:   &  ~~5.8:    & ~74:   & Chen \etal 1999 \\
HDFN--JD1    &  12.5 ?? & 23.9  & -23.9  &   13.7     & 175    & Dickinson \etal 2000 \\
\hline
\end{tabular}
\end{table}

CDFa--G1 at $z = 4.815$ is the most luminous Lyman break galaxy yet 
identified (among both $U_n$-- and $G$--dropouts) in our large, ground--based 
survey.  The spectroscopically verified $5 < z < 6$ galaxies have luminosities 
that are mostly fairly typical of $z \approx 3$ LBGs, ranging from 0.66 to 
1.5$\times L^\ast$.  The $z = 5.74$ object from Hu, Cowie \& McMahon (1999)
has been described as `extremely luminous,' but is actually quite typical 
for $z \approx 3$ to 4 LBGs, and somewhat fainter than HDF~3--951 at 
$z=5.33$.   Chen \etal (1999) have identified a candidate galaxy 
at $z = 6.68$ from STIS slitless spectra.  The available photometry is limited, 
but based on the spectral continuum flux density estimate, this object appears to 
be significantly more luminous than other known LBGs at $2 < z < 6$, or than any
$0 < z < 10$ candidates from the HDF/NICMOS sample.   This is a remarkable result, 
if true, since the solid angle covered by the STIS field is $< 1$~arcmin$^2$.  
On the other hand, if the $J$--dropout object HDFN--JD1 were really at 
$z = 12.5$, it would be more luminous still, nearly 3$\times$ brighter 
than CDFa--G1 at $z=4.82$, with a `raw' UV SFR = 175 $M_\odot$~yr$^{-1}$.

\section{Discussion}

Overall, the HDF/NICMOS data demonstrate both the promise and the
challenges which lie ahead for finding and studying the `first'
galaxies.   The rest--frame optical view of Lyman break galaxies 
presented in \S\S2 and 3 strongly suggests that the galaxy population 
at $2 < z < 3$ had not yet achieved maturity.  The giant, Hubble 
sequence spirals and ellipticals that dominate the high--mass end 
of the galaxy population today are not seen at $z > 2$. 
In a sample of HDF galaxies selected in the near--infrared, 
nearly all galaxies with spectroscopic or plausible photometric 
redshifts $2 < z < 3.5$ are evidently forming stars quite rapidly 
and can also be identified via their emitted--frame UV light.
The evidence from SCUBA shows that there are occasional `monsters' 
whose obscured star formation may be quite important to the global 
emissive energy budget from galaxies.  The identification of these 
objects, relatively rare by number, remains an important dilemma.

Broad band color selection has been the most successful means for 
identifying high redshift galaxies, but we seem to be pushing the 
limits of what can be accomplished at $z > 6$ with present--day
capabilities.  The NICMOS HDF images are the deepest near--IR data 
now available, and they do include plausible candidates for galaxies 
at $6 < z < 9$, but they are relatively few, and most are quite 
probably too faint for spectroscopic confirmation.  We probably 
should not expect to find galaxies much {\it brighter} than these 
candidates unless some of the `first' galaxies were significantly 
more luminous than the boring old `later' galaxies that we have 
now surveyed extensively at $z \approx 3$.  This is not impossible
of course:  the Chen \etal object and HDFN--JD1 are both possible 
(but unconfirmed) $z > 6$ candidates more luminous than any 
normal Lyman break galaxy at $z < 5$.  Perhaps indeed there 
are very luminous, relatively unobscured proto--galaxies
out there waiting to be found,  a hope that was once quite 
widespread, but which seems to have gradually faded in the 
modern era of 25th magnitude LBGs and optically invisible
SCUBA sources.  Perhaps it will still make a comeback...

These few rather speculative candidates aside, the evidence from
the HDF alone would suggest that the population of UV--bright LBGs 
may be thinning out at $z > 5$, at least for objects comparable to
those at the bright end of the $z \approx 3$ luminosity function.
It should be remembered, however, that this was also the conclusion
reached by Madau \etal (1996) at $z > 3.5$, a result that has since
been challenged by larger surveys with extensive spectroscopy.  
It is undoubtedly dangerous to draw conclusions too strongly 
from one 5~arcmin$^2$ field.  However, extending this work 
to larger areas and more sightlines will be an expensive effort.
Surface brightness dimming and limited solid angle coverage may 
limit our ability to see much more with NICMOS (assuming that it
is successfully revived in 2001), and ground--based near--IR imaging 
may never go deep enough to detect any but the most luminous objects
at $z > 5$.   Wider fields imaged with the {\it HST} WFC3 near--IR
channel (coming circa 2004) may offer the best survey opportunity until 
NGST, but a substantial investment of observing time will be needed 
to survey adequate solid angles to sufficient depth.  

Alternatively, we may turn to other observing strategies, e.g., 
by taking advantage of gravitational lensing from foreground galaxy 
clusters to boost very distant objects to detectable magnitudes.
Narrow band and blind multislit emission line searches are being 
carried out through airglow windows (e.g., at $\lambda 9150$\AA, 
corresponding to $z \approx 6.5$;  cf.\ Crampton \& Lilly 1999; 
Stockton 1999).  Or perhaps concerted efforts to identify SCUBA 
sources will indeed turn up objects at $z \gg 5$, where the 
advantage of the negative sub--mm $k$--correction is enormous.

These data are offering a first glimpse into the so--called 
`dark ages,' and giving hope that there may be luminous things 
there to find and study.  In some sense, we may not know that 
we've found the first galaxies until we can find no more 
beyond them.  Holding to that standard will ensure that 
a more challenging (and hence, more rewarding) threshold of proof 
always lies ahead.

\begin{acknowledgements}

I thank my HDF/NICMOS collaborators for their contributions and 
for permission to show results in advance of publication, particularly 
Casey Papovich, Tamas Budav\'ari, Jean--Marc Deltorn, Chris Hanley 
and Harry Ferguson.   I also thank Chuck Steidel, Max Pettini, 
Mauro Giavalisco, Kurt Adelberger, and Alice Shapely for their 
long--standing and highly valued collaboration.  I am grateful to 
the Royal Society for inviting me to this meeting and for supporting 
my travel and accommodations, and to the editors of the proceedings for 
their patience.  This work was supported by NASA grant GO-07817.01-96A.

\end{acknowledgements}

\label{lastpage}

\end{document}